\documentclass[%
aps,prb,superscriptaddress,twocolumn,floatfix,10pt,longbibliography]{revtex4-2}
\usepackage[utf8]{inputenc}
\usepackage{amsmath,amssymb}
\usepackage{empheq}
\usepackage{lipsum}
\usepackage{mathtools,cuted}
\usepackage{esvect}
\usepackage{multirow}
\usepackage{xcolor}
\usepackage{soul}
\usepackage[export]{adjustbox}
\usepackage{graphicx}
\usepackage{dcolumn}
\usepackage{bm}
\usepackage{hyperref}
\usepackage{nameref}
\usepackage{physics}
\usepackage[mathlines]{lineno}
\usepackage[capitalise]{cleveref}
\usepackage{bbm}
\usepackage{orcidlink}
\newcommand{\ham}[0]{H}
\newcommand{\hamOP}[0]{\hat{\ham}}

\newcommand{\en}[0]{E}
\newcommand{\sbb}{\overline{SS}}
\newcommand{\ssb}{S\sbar}
\newcommand{\sbs}{\sbar{}S}
\newcommand{\sbar}{\overline{S}}
\newcommand{\ISR}[2]{\widetilde{\ham}_{#1}(#2)}

\begin{document}
	
\title{Latent Haldane Models}

\author{Anouar Moustaj\,\orcidlink{0000-0002-9844-2987}}
 \affiliation{Institute of Theoretical Physics, Utrecht University, Utrecht, 3584 CC, Netherlands}
\author{Lumen Eek\,\orcidlink{0009-0009-1233-4378}}
 \affiliation{Institute of Theoretical Physics, Utrecht University, Utrecht, 3584 CC, Netherlands}
 \author{Malte Röntgen\,\orcidlink{0000-0001-7784-8104}}
 \affiliation{
 	Laboratoire d’Acoustique de l’Université du Mans, Unite Mixte de Recherche 6613, Centre National de la Recherche Scientifique, Avenue O. Messiaen, F-72085 Le Mans Cedex 9, France
 }
\author{Cristiane Morais Smith\,\orcidlink{0000-0002-4190-3893}}
\affiliation{Institute of Theoretical Physics, Utrecht University, Utrecht, 3584 CC, Netherlands}

	\date{\today}
	
\begin{abstract}
Latent symmetries, which materialize after performing isospectral reductions, have recently been shown to be instrumental in revealing novel topological phases in one-dimensional systems, among many other applications. In this work, we explore how to construct a family of seemingly complicated two-dimensional models that result in energy-dependent Haldane models upon performing an isospectral reduction. In these models, we find energy-dependent latent Semenoff masses without introducing a staggered on-site potential. In addition, energy-dependent latent Haldane masses also emerge in decorated lattices with nearest-neighbor complex hoppings. Using the Haldane model's properties, we then predict the location of the topological gaps in the aforementioned family of models and construct phase diagrams to determine where the topological phases lie in parameter space. This idea yielded, for instance,
useful insights in the case of a modified version of $\alpha$-graphyne and hexagonal plaquettes with additional decorations, where the gap-closing energies can be calculated using the ISR to predict topological phase transitions.
\end{abstract}
	
	\maketitle{}
	
	
\section{Introduction}\label{sec:Intro}

Topological phases of matter have become a cornerstone of modern condensed matter theory, with applications ranging from material science and photonics to mechanical and non-Hermitian systems \cite{Hasan2010ColloquiumInsulators,Weimann2017TopologicallyCrystals,Kane2014TopologicalLattices,Ma2019TopologicalSystems,Gong2018TopologicalSystems}. 
A well-established theoretical framework that characterizes topological phases has been developed and refined in the last three decades, based on a classification of disordered insulators and superconductors subject to anti-unitary symmetries \cite{Altland1997NonstandardStructures,Schnyder2008ClassificationDimensions,Ryu2010TopologicalHierarchy,Ludwig2016TopologicalBeyond}.
These phases have been uncovered in a variety of systems, from periodic crystals to amorphous, aperiodic, fractals, and disordered systems \cite{Agarwala2017TopologicalSystems,Moustaj2025AnomalousInsulators,Osseweijer2024HaldaneGasket,Grushin2022TopologicalMatter}. One of the main properties of topological insulators is the bulk-boundary correspondence, which implies the existence of robust in-gap edge modes in one dimension (1D) and \textit{gapless} boundary states in two and three dimensions (2D and 3D). 

A paradigmatic model of topological insulators in 2D is the Haldane model \cite{Haldane1988ModelAnomaly}. It models tightly bound spinless electrons on a 2D honeycomb lattice subject to time-reversal (TR) symmetry breaking by including complex next-nearest-neighbor hoppings. One of the consequences of nontrivial topology in such a system is the nonexistence of a complete set of exponentially localized Wannier functions, also known as an \textit{obstruction to Wannierization} \cite{Brouder2007ExponentialInsulators}. Consequently, a topologically nontrivial state cannot be adiabatically connected to an atomic limit in which electrons occupy maximally localized Wannier states \cite{Andrews2024LocalizationSystems}. This model forms the backbone for understanding the quantum spin Hall effect \cite{Kane2005QuantumGraphene,Kane2005Z2Effect}, and is also the simplest model to realize the quantum anomalous Hall effect \cite{Chang2023Colloquium:Effect}. Because of its importance, it has been dubbed the ``hydrogen atom'' of topological insulators \cite{Bernevig2013SimpleInsulator}. Here, we will explore how it can be combined with the isospectral reduction (ISR) technique \cite{Bunimovich2014IsospectralTransformationsNewApproach}. 

In the last few years, the ISR has been used to understand and uncover many phenomena that were previously difficult to grasp. This technique originates in the study of large networks using graph theory, where the adjacency matrices for complicated graphs are reduced on a set of chosen vertices using a projective scheme. As the name implies, the technique reduces the matrix dimension while preserving the spectrum at the cost of introducing a non-linear eigenvalue problem. Initially, the method was shown to be very useful in revealing hidden network structures \cite{Bunimovich2019AMNS4231FindingHiddenStructuresHierarchies} and obtaining better eigenvalue approximations \cite{Bunimovich2012LAIA4371429IsospectralGraphReductionsImproved}. In the context of physics, it takes the form of an effective Hamiltonian. It has found use in various applications, including the unveiling of latent symmetries \cite{Smith2019PA514855HiddenSymmetriesRealTheoretical}, explaining accidental degeneracies \cite{Rontgen2021PRL126180601LatentSymmetryInducedDegeneracies}, designing quantum information transfer protocols \cite{Rontgen2020PRA101042304DesigningPrettyGoodState}, or uncovering novel topological phases of Hermitian and non-Hermitian Hamiltonians \cite{Eek2024EmergentModels,Rontgen2024TopologicalSymmetry,Eek2024Higher-orderSymmetries}. 

In this work, we use ISRs in a range of 2D lattice models to understand some of their features in terms of the Haldane model. In particular, we study a set of selected systems that, upon applying an ISR, reduce to an energy-dependent Haldane model. In doing so, we uncover the mechanisms behind the formation of gaps by breaking latent symmetries, that is, hidden symmetries of the system that become obvious only in the ISR. Specifically, we see that a latent Semenoff mass is generated whenever the parameters of the model destroy a latent inversion symmetry. These gaps appear at energies that can be predicted based on the latent Haldane model. Furthermore, when TR symmetry is broken by introducing complex \textit{nearest-neighbor} hoppings, it is also possible to find topological phases at fillings dictated by specific energy conditions, and directly construct topological phase diagrams from the latent Haldane model. This is in contrast to the usual Haldane model, in which the topological phase is induced by next-nearest-neighbor hopping. As an example, we apply this idea to a modified version of $\alpha$-graphyne \cite{Li2014GraphdiyneConstruction,vanMiert2014Tight-bindinggraphynes} and a decorated hexagonal plaquette. We predict the gap-closing energies and critical parameter values using the ISR. We emphasize that these features are not model-specific but hold for a family of models that can be constructed from the simple building blocks that we thoroughly study in this paper.  

This article is structured as follows: in \cref{sec: real Haldane}, for pedagogical reasons, we introduce the Haldane model and recall its most essential features, such as its topological phases and the associated edge states in a ribbon geometry. We explain how the phase diagram can be constructed using a low-energy approximation near the gap-closing point and introduce the terminology used in the rest of the manuscript. This is followed by an explanation of the ISR in \cref{sec: ISR} and an application to $\alpha$-graphyne in \cref{Sec: proof of principle}. We calculate the gap-closing energies and introduce the ingredients necessary to generate topological phases. Then, we present the first system in which a latent Semenoff mass is generated from a decorated lattice in \cref{Sec: Latent Semenoff}. It provides the realization of a gap-opening mechanism that takes the form of a Semenoff mass, without a staggered on-site potential. We also introduce complex long-range hoppings that yield topological phases and construct a topological phase diagram. In \cref{Sec: Latent Hald mass}, we consider a system that now generates a latent Haldane mass from complex nearest-neighbor hoppings. When combined with the decorated lattice generating a latent Semenoff mass, this yields a variety of topological phases, for which we determine the phase diagram for the fillings that admit them. 
Finally, in \cref{Sec: Conclusion}, we summarize our findings and discuss potential future directions of study.

\section{Haldane model}\label{sec: real Haldane}

We start by reviewing the physics of the Haldane model on a honeycomb lattice, one of the earliest Chern insulators, exhibiting the quantum anomalous Hall effect \cite{Haldane1988ModelAnomaly}. It is a tight-binding model for a spinless electron hopping in a honeycomb lattice only through nearest neighbors. The anomalous Hall effect is then implemented through the addition of the Haldane mass, that is, a TR symmetry-breaking term, which is incorporated by next-nearest neighbor complex hoppings, for which the phase is determined by the directionality (i.e., clockwise or not) of the path between the sites. Finally, the Haldane model also contains an inversion symmetry-breaking term called the Semenoff mass \cite{Semenoff1984Condensed-MatterAnomaly}.

\begin{figure}[!hbt]
    \centering
    \includegraphics[width=\columnwidth]{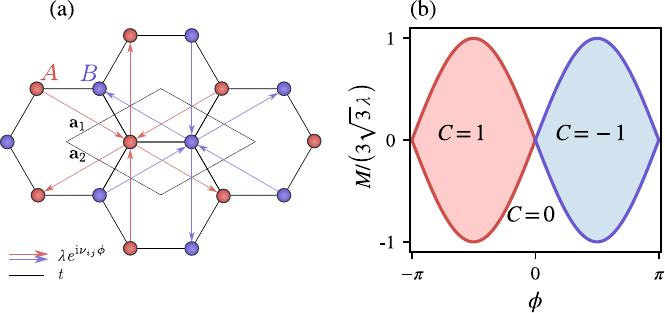}
        \caption{(a) Sketch of the Haldane Model. The arrows on the next-nearest-neighbor couplings determine the sign of the Haldane phase. (b) Phase diagram of the Haldane model in terms of the Chern number $C$.}
        \label{fig:HaldaneModel}
\end{figure}


 The system is sketched in \cref{fig:HaldaneModel}(a), and the Hamiltonian for this model reads  
\begin{equation}
    \hat{H} \equiv t\sum_{\langle i,j\rangle}c^\dagger_ic_j+\lambda\sum_{\langle\langle i,j\rangle\rangle}e^{i\nu_{ij}\phi}c^\dagger_ic_j+M\sum_j(-1)^{\mu_j}c^\dagger_jc_j,
\end{equation}
where $t$ is the nearest-neighbour hopping parameter, $\lambda$ is the next-nearest neighbour hopping strength, and $M$ is the Semenoff mass. Additionally,
\begin{equation*}                   
    \nu_{ij}=\text{sign}\left[(\mathbf{d}_{im}\times\mathbf{d}_{mj})\cdot\hat{\mathbf{z}}\right]=\pm1,
\end{equation*}
where $\mathbf{d}_{im}$ is the vector pointing from $i$ to its nearest-neighbor $m$, and the phase $\phi$ is associated with the internal flux experienced by the electron along the path. Finally, $\mu_j=0$ ($1$), depending on whether the site belongs to the $A$ ($B$) sublattice. Working out the Bloch Hamiltonian, we obtain
\begin{equation}\label{Eq: Haldane Bloch Hamiltonian}
    H(\mathbf{k})=\begin{pmatrix}
        M+\lambda f_{\phi}(\mathbf{k}) & tg(\mathbf{k}) \\
        tg^*(\mathbf{k})& -M+\lambda f_{-\phi}(\mathbf{k})
    \end{pmatrix}
\end{equation}
where 
\begin{align*}
    g(\mathbf{k})& = \left(1+e^{i\mathbf{k}\cdot\mathbf{a}_1}+e^{i\mathbf{k}\cdot\mathbf{a}_2}\right) \\
    f_\phi(\mathbf{k}) &= \begin{multlined}[t]
        2\bigg\{\cos\left (\mathbf{k}\cdot\mathbf{a}_1-\phi\right)+ \\ \cos\left (\mathbf{k}\cdot\mathbf{a}_2+\phi\right)+\cos\left [\mathbf{k}\cdot(\mathbf{a}_1-\mathbf{a}_2)+\phi\right]\bigg\},
    \end{multlined}
\end{align*}
and $\mathbf{a}_1$, $\mathbf{a}_2$ are the lattice vectors shown in \cref{fig:HaldaneModel}(a).

Using the Bloch Hamiltonian \cref{Eq: Haldane Bloch Hamiltonian}, one can calculate the Chern number 
\begin{equation}\label{Eq: Chern number}
    C = \int_{\text{B.Z.}}\frac{\dd\mathbf{k}}{2\pi}F_{xy}(\mathbf{k}),
\end{equation}
where $F_{xy}(\mathbf{k})$ is the Berry curvature, 
\begin{equation*}
    F_{xy}(\mathbf{k})=-i\left[\bra{\frac{d\psi_0(\mathbf{k})}{dk_x}}\ket{\frac{d\psi_0(\mathbf{k})}{dk_y}}-\bra{\frac{d\psi_0(\mathbf{k})}{dk_y}}\ket{\frac{d\psi_0(\mathbf{k})}{dk_x}}\right].
\end{equation*}
Here, $\psi_0(\mathbf{k})$ is the Bloch state of the lowest band. The Chern number can be interpreted through the TKNN formula \cite{Thouless1982PRL49405QuantizedHallConductanceTwoDimensional} as the quantized Hall conductivity $\sigma_{xy}$ (in units of $e^2/h$).
A phase diagram of the Chern number, as a function of the Semenoff mass $M$ and Haldane phase $\phi$, is shown in \cref{fig:HaldaneModel}(b).

\begin{figure}[!hbt]
    \centering
    \includegraphics[width=\columnwidth]{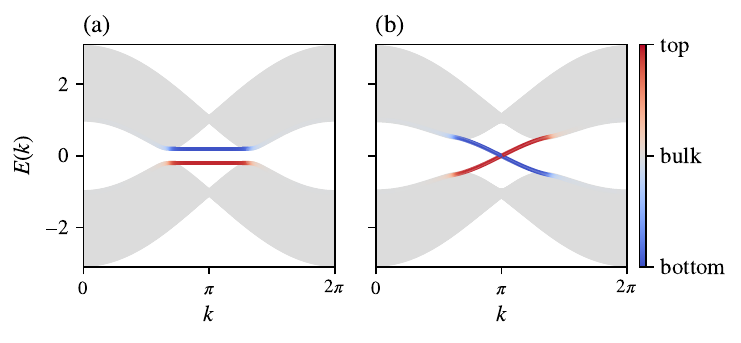}
        \caption{Band structure of the Haldane model for a ribbon geometry. (a) With only a Semenoff mass, the model is a trivial insulator, with nondispersive edge modes. The parameters used are $(M,\lambda) = (0.1t, 0)$. (b) The Haldane mass makes the model a topological insulator, with dispersive edge modes crossing the bulk band. The value of the parameters are $(M,\lambda) = (0, 0.1t)$.}
    \label{fig:HaldaneEdgeModes}
\end{figure}

To better understand the phase diagram, let us briefly discuss the band structure of \cref{Eq: Haldane Bloch Hamiltonian}.
In general, this band structure is gapped.
However, when the two mass terms fulfill
\begin{equation} \label{eq:gapClosingMasses}
    M=\pm3\sqrt{3}\lambda \sin\phi
\end{equation}
(which is, in particular, fulfilled if both $M$ and $\lambda$ vanish), the spectrum is gapless; the two bands cross at 
\begin{equation}\label{eq:gapClosingLocationMinus}
 (\mathbf{k},E) = (\mathbf{K},-3 \lambda \cos\phi)
\end{equation}
for $M=-3\sqrt{3}\lambda \sin\phi$, and at
\begin{equation} \label{eq:gapClosingLocationPlus}
 (\mathbf{k},E) = (\mathbf{K}',-3 \lambda \cos\phi)
\end{equation}
for $M=+3\sqrt{3}\lambda \sin\phi$.
$\mathbf{K}$ and $\mathbf{K}'$ are high-symmetry points satisfying $\mathbf{K}\cdot\mathbf{a}_i= 2\pi/3,\ 4\pi/3$ and $\mathbf{K}'\cdot\mathbf{a}_i=4\pi/3,\ 2\pi/3$, for $i=1,2$ respectively.

As discussed above, when both $\lambda$ and $M$ are zero, the system described by the Hamiltonian \cref{Eq: Haldane Bloch Hamiltonian} is gapless.
The gap-closing points, i.e., the Dirac points, are situated at $E=0$ and are globally stable as long as the system possesses $C_3$ symmetry.
A gap opens when inversion symmetry is broken by a Semenoff mass $M\neq0$, or when breaking TR symmetry with the Haldane mass $\lambda\neq0$.
Note that the gap closes again when both mass terms are non-vanishing and additionally fulfill \cref{eq:gapClosingMasses}.

When only inversion symmetry is broken, the system becomes a trivial insulator at half-filling. This can be observed from the open-geometry spectrum in \cref{fig:HaldaneEdgeModes}(a), where the edge modes do not traverse the gap, or by computing the Hall conductivity $\sigma_{xy}$, which is 0. However, when adding the TR symmetry-breaking Haldane mass, the system may become a topological insulator.
In contrast with the trivial gap, the edge modes \textit{must} traverse the gap to connect the two Dirac points [see \cref{fig:HaldaneEdgeModes}(b)]. 
This is an example of bulk-boundary correspondence, where a property of the bulk (the Chern number) predicts the presence of gap-traversing, dispersive modes at the edge. 

The phase diagram in \cref{fig:HaldaneModel}(b) can be constructed by analyzing the continuum Dirac Hamiltonian near the high-symmetry points $K$ and $K'$. One can algebraically manipulate \cref{Eq: Haldane Bloch Hamiltonian} into
\begin{equation*}
    H(\mathbf{k})=\epsilon(\mathbf{k})\mathbbm{1}_{2\times2}+\sum_{i=1}^3d_j(\mathbf{k})\sigma_j,
\end{equation*}
where 
\begin{widetext}
    \begin{equation}\label{Eq: d vector Haldane}
    \begin{split}
        d_1(\mathbf{k})&= t\left[\cos\left(\mathbf{k}\cdot\mathbf{a}_1\right)+\cos\left(\mathbf{k}\cdot\mathbf{a}_2\right)+1\right], \\
        d_2(\mathbf{k})&= t\left[\sin\left(\mathbf{k}\cdot\mathbf{a}_1\right)+\sin\left(\mathbf{k}\cdot\mathbf{a}_2\right)\right], \\
        d_3(\mathbf{k})&=M+2\lambda\sin\phi\bigg\{\sin\left(\mathbf{k}\cdot\mathbf{a}_1\right)-\sin\left(\mathbf{k}\cdot\mathbf{a}_2\right)-\sin\left[\mathbf{k}\cdot(\mathbf{a}_1-\mathbf{a}_2)\right]\bigg\}, \\
        \epsilon(\mathbf{k})&=2\lambda\cos\phi\bigg\{\cos\left(\mathbf{k}\cdot\mathbf{a}_1\right)+\cos\left(\mathbf{k}\cdot\mathbf{a}_2\right)+\cos\left[\mathbf{k}\cdot(\mathbf{a}_1-\mathbf{a}_2)\right]\bigg\}. \\
    \end{split}
\end{equation}
\end{widetext}

After a rather long calculation \cite{Bernevig2013TopologicalSuperconductors}, where one expands the momentum near $\mathbf{K}$ and $\mathbf{K}'$ up to first order, one obtains the following continuum theory near the Dirac points
\begin{widetext}
  \begin{equation}\label{Eq:DiracHamiltonians}
    \begin{split}
       h(\mathbf{K}+\mathbf{k})&\approx-3\lambda\cos\phi\mathbbm{1}_{2\times2}+\frac{3}{2}t\left(k_2\sigma_1-k_1\sigma_2\right)+\left(M+3\sqrt{3}\lambda  \sin\phi\right)\sigma_3, \\
    h(\mathbf{K}'+\mathbf{k})&\approx-3\lambda\cos\phi\mathbbm{1}_{2\times2}-\frac{3}{2}t\left(k_2\sigma_1+k_1\sigma_2\right)+\left(M-3\sqrt{3}\lambda  \sin\phi\right)\sigma_3, 
    \end{split}
\end{equation}  
\end{widetext}
with $\mathbf{k}=(k_1,k_2)$ and $||\mathbf{k}||<<||\mathbf{K}||$. 
The Chern number in \cref{Eq: Chern number} can be analytically calculated in this case, yielding
\begin{equation} \label{eq:continuumHallConductivity}
    C=\frac{1}{2}\text{sign}\left(M-3\sqrt{3}\lambda \sin\phi\right)\text{sign}(\det\mathcal{A}),
\end{equation}
where the matrix $\mathcal{A}$ is defined such that the Dirac Hamiltonian in \cref{Eq:DiracHamiltonians} is written as:
\begin{equation*}
    h(\mathbf{k})=\sum_{i,j=1}^2k_i\mathcal{A}_{ij}\sigma_j+\mathcal{M}\sigma_z,
\end{equation*}
 and $\mathcal{M}$ is the momentum-independent factor multiplying $\sigma_z$. In this case, $\mathcal{M}=M-3\sqrt{3}\lambda \sin\phi$.  Equation \ref{eq:continuumHallConductivity} indicates that in the continuum theory, the Chern number is quantized to a half-integer and does not describe the lattice Hall conductivity completely. However, it is valuable for computing \textit{changes} in the conductivity, which are still integer-valued. 

The lattice conductivity is zero if one starts from a trivial atomic limit, where $M\to\infty$ and all sites are effectively decoupled. Lowering $M$ all the way down to $3\sqrt{3}\lambda \sin\phi$ (assuming $\phi>0$), we encounter the first gap closing at the $K'$ point [cf. \cref{eq:gapClosingLocationPlus}]. The Chern number changes from 1/2 to -1/2, meaning the system is now in a topological phase with $\sigma_{xy}=-1$. We continue lowering $M$ further down until $M=-3\sqrt{3}\lambda \sin\phi$, where the gap closes at the $K$ point. The conductivity changes from -1/2 back to 1/2, signaling that we are once again in the $\sigma_{xy}=0$ phase. Doing the same analysis for $\phi<0$, we obtain the phase diagram shown in \cref{fig:HaldaneModel}(b). A more detailed textbook analysis of the Haldane model can be found in Ref.~\cite{Bernevig2013TopologicalSuperconductors}.

\section{Latent Haldane Models}\label{sec: Latent Haldane models}
After the preliminary discussions above, we now introduce the "latent Haldane" models.
Before doing so, we provide a brief introduction of the underlying tool used, namely, the ISR.

\subsection{Isospectral reduction}\label{sec: ISR}

The ISR, which is akin to an effective Hamiltonian, is given by
\begin{equation} \label{eq:ISRDef}
    \ISR{S}{\en} = \ham_{SS} - \ham_{\ssb} \left(\ham_{\sbb} - \en{}\, I \right)^{-1} \ham_{\sbs} \,,
\end{equation}
where $\ham$ is the matrix form of the Hamitonian $\hat{H}$ in the basis of single-site excited states.
In the following, when speaking of site $i$, we mean the basis state for which only site $i$ is excited.
In \cref{eq:ISRDef}, $S$ denotes a set of sites over which we reduce, and $\sbar$ denotes its complement, that is, the other sites. $I$ denotes the identity matrix, which has the same dimension as $\ham_{\sbb}$.
The sub-matrices $\ham_{XY}$ are obtained by taking only the rows $X$ and the columns $Y$ from the full matrix $\ham$.
We note that the ISR can be derived as follows.
One starts by writing the original matrix eigenvalue problem $\ham \mathbf{\Psi} = E \mathbf{\Psi}$ (with $\mathbf{\Psi}$ the eigenvectors of the Hamiltonian $\ham$) in block form as \footnote{We note that one might have to change the numbering of sites to obtain this specific block form; that is, enumerating the sites such that the first $|S|$ sites are those in $S$, and the following are those in $\sbar$, with $|S|$ denoting the number of sites in the set $S$. We further note that such a change of the enumeration of the sites corresponds to applying a similarity transformation $P^{-1} \ham P$ to $\ham$, with $P$ a permutation matrix.}
\begin{equation} \label{eq:isrEq}
    \begin{pmatrix}
        \ham_{SS} & \ham_{\ssb} \\
        \ham_{\sbs} & \ham_{\sbb}
    \end{pmatrix}
    \begin{pmatrix}
        \mathbf{\Psi}_{S} \\
        \mathbf{\Psi}_{\sbar}
    \end{pmatrix}
    = E     \begin{pmatrix}
        \mathbf{\Psi}_{S} \\
        \mathbf{\Psi}_{\sbar}
    \end{pmatrix}
\end{equation}
where $\mathbf{\Psi}_{X}$ denotes the vector obtained from $\mathbf{\Psi}$ by taking only the components on $X$.
Multiplying out \cref{eq:isrEq} yields two coupled equations \footnote{Namely, $\ham_{SS} \mathbf{\Psi}_{S} + \ham_{\ssb} \mathbf{\Psi}_{\sbar} = E \mathbf{\Psi}_{S}$ and $\ham_{\sbs} \mathbf{\Psi}_{S} + \ham_{\sbb} \mathbf{\Psi}_{\sbar} = E \mathbf{\Psi}_{\sbar}$}; solving the second for $\mathbf{\Psi}_{\sbar}$ and inserting it into the first yields the non-linear eigenvalue problem
\begin{equation} \label{eq:nonLinearEVP}
    \ISR{S}{\en} \mathbf{\Psi}_{S} = E \mathbf{\Psi}_{S} \,,
\end{equation}
with $\ISR{S}{\en}$ the ISR.

The name of the ISR stems from the fact that, under very mild conditions on $\ham$, the eigenvalues of $\ISR{S}{\en}$ \footnote{As can be deduced from \cref{eq:nonLinearEVP}, the eigenvalues of $\protect\ISR{S}{\en}$ are the values of $\en$ for which $Det(\protect\ISR{S}{\en} - \en I) = 0$.} are exactly the eigenvalues of the original Hamiltonian $\ham$; that is, $\ham$ and $\ISR{S}{\en}$ are isospectral \cite{Bunimovich2014IsospectralTransformationsNewApproach}.

Before we continue, let us remark that the ISR has been used in the past few years in different areas.
A non-exhaustive list of topics and articles articles employing the ISR comprises several graph-theoretical problems \cite{Bunimovich2014IsospectralTransformationsNewApproach,Kempton2020LAIA594226CharacterizingCospectralVerticesIsospectral,Bunimovich2011N25211IsospectralGraphTransformationsSpectral, Bunimovich2012C22033118IsospectralCompressionOtherUseful, Bunimovich2012LAIA4371429IsospectralGraphReductionsImproved,VasquezFernandoGuevara2014NLAA22145PseudospectraIsospectrallyReducedMatrices,Smith2019PA514855HiddenSymmetriesRealTheoretical,Kempton2024AC7225IsospectralReductionsQuantumWalks,duarteIsospectralReductionInfinite2020,baravieraIsospectralReductionsNonnegative2023}, crystals \cite{Eek2024PRB109045122EmergentNonHermitianModels, Rontgen2021PRL126180601LatentSymmetryInducedDegeneracies, Rontgen2024PRB110035106TopologicalStatesProtectedHidden}, fractals \cite{Kempkes2019DesignGeometry}, waveguide networks
\cite{Rontgen2023PRL130077201HiddenSymmetriesAcousticWave}, non-Hermitian \cite{Cui2023PRL131237201ExperimentalRealizationStableExceptional} and non-linear systems \cite{fangExceptionalFeaturesNonlinear2024}, granular setups  \cite{Zheng2023PRB108L220303RobustTopologicalEdgeStates} or intelligent surfaces \cite{prodhommeEfficientComputationPhysicsCompliant2023}.

\subsection{Proof of principle: $\alpha$-graphyne} \label{Sec: proof of principle}
\begin{figure}[!hbt]
    \centering
    \includegraphics[]{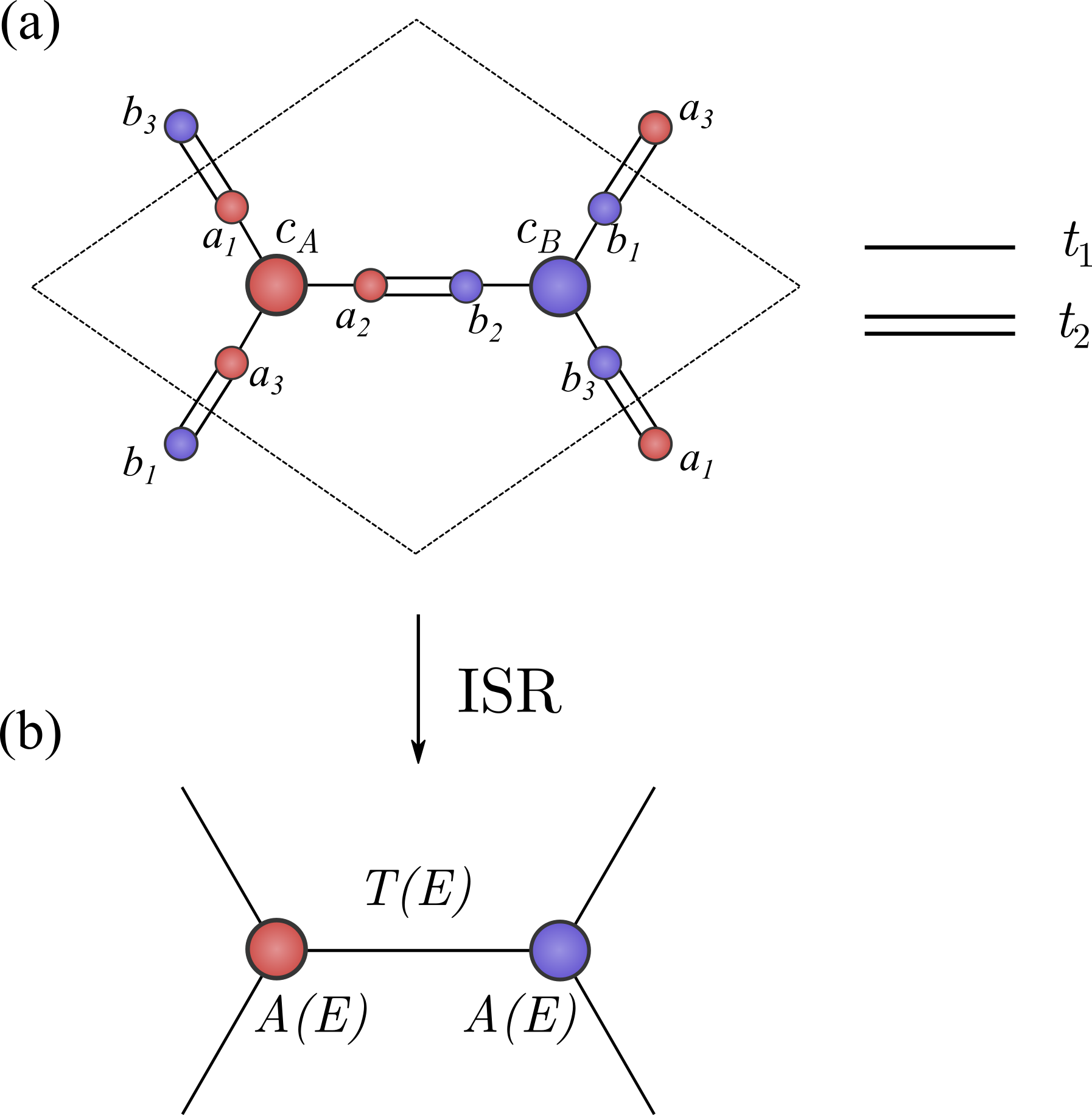}
    \caption{(a) Unit cell of the $\alpha$-graphyne lattice and (b) ISR to the regular graphene unit cell. The ISR is performed on sites $A$ and $B$, marked by larger circles than the rest in (a).}
    \label{fig: graphyne lattice}
\end{figure}

To keep things simple, we start by analyzing spinless $\alpha$-graphyne \cite{Li2014GraphdiyneConstruction,vanMiert2014Tight-bindinggraphynes}, shown in \cref{fig: graphyne lattice}(a).
The model is described by the following Hamiltonian in real space, 
\begin{equation}
    \begin{split}
    \hamOP =& t_1\sum_{i}\sum_{\mu=1}^3\left[c^\dagger_{i,A}a_{i,\mu} + c^\dagger_{i,B} b_{i,\mu}\right] + t_2\sum_{i} a^\dagger_{i,2} b_{i,2} \\
    &+t_2\sum_{\langle i,j\rangle}(a^\dagger_{i,1}b_{j,1}+a^\dagger_{i,3}b_{j,3})+\text{h.c.}        
    \end{split}
\end{equation}
where the operators create (annihilate) spinless electrons on sites following the labels shown in \cref{fig: graphyne lattice}(a), and $i,j$ are cell indices. This model contains eight sites per unit cell, making it an eight-band model. 

Performing an ISR onto the sites $A$ and $B$ in each unit cell, as depicted in \cref{fig: graphyne lattice}(b), and taking the Bloch-Hamiltonian of the resulting lattice yields the $2\times 2$ energy-dependent Bloch Hamiltonian 
   \begin{equation}\label{Eq: ISR first}
    \Tilde{H}_E(\mathbf{k})=\begin{pmatrix}
        A(E) & T(E) g(\mathbf{k}) \\
        T(E) g^*(\mathbf{k}) & A(E)
    \end{pmatrix},
\end{equation} 
where
\begin{align}
    A(E) &= \frac{3Et_1^2}{E^2-t_2^2}, \notag\\
    T(E) &= \frac{t^2_1t_2}{E^2-t_2^2}.
\end{align}

As a warm up, let us now discuss how the energy-dependent terms $A(E)$ and $T(E)$ influence the behavior of the model.
Firstly, we realize that the eigenvalue equation of \cref{Eq: ISR first}, $\Tilde{H}_E(\mathbf{k})\Psi_S(\mathbf{k})=E\Psi_S(\mathbf{k})$, can be rewritten as 
\begin{equation}
\begin{pmatrix}
        0 & T(E) g(\mathbf{k}) \\
        T(E) g^*(\mathbf{k}) & 0
    \end{pmatrix} \Psi_S(\mathbf{k}) = \epsilon \Psi_S(\mathbf{k})
\end{equation}
with $\epsilon = E - A(E)$.
This is exactly the eigenvalue equation of the conventional Haldane model, \cref{Eq: Haldane Bloch Hamiltonian}, with $M=\lambda = 0$ (both Haldane and Semenoff mass set to zero), energy-dependent coupling $T(E)$ and eigenvalue $\epsilon(E)$.
And since the gap between the two bands of the conventional Haldane model closes at $E=0$\footnote{With $\mathbf{k}$ being equal to either $\mathbf{K}$ or $\mathbf{K}'$, for which $g(\mathbf{k})$ vanishes.}, we see that the gap(s) in our latent Haldane model close for energies fulfilling $\epsilon(E) = E - A(E)=0$.


Having discussed the latent Haldane model with both Semenoff and Haldane mass set to zero, let us now introduce these masses to the model.
A simple way starts by realizing that any modification of the form $\ham_{SS} \rightarrow \ham_{SS} + X$ modifies the isospectral reduction as $\ISR{S}{\en} \rightarrow \ISR{S}{\en} + X$.
In other words, \cref{eq:ISRDef} tells us that any modification of the original Hamiltonian that concerns only the sites $S$ leads to a corresponding \emph{energy-independent} modification of the isospectral reduction.
Thus, we can modify the original Hamiltonian by adding a Semenoff mass $M$ to the $A$ and $B$ sites, and also include complex hoppings between $A$ sites of neighboring unit cells and $B$ sites of neighboring unit cells with hopping strength $\lambda$, resulting in
   \begin{equation*}
    H_E(\mathbf{k})=\Tilde{H}_E(\mathbf{k}) + \begin{pmatrix}
        M+\lambda f_{\phi}(\mathbf{k}) & 0 \\
        0 & -M+\lambda f_{-\phi}(\mathbf{k})
    \end{pmatrix} \,.
\end{equation*}
This implies that the same methodology for analyzing topological phase transitions applied to the original Haldane model can be extended to this scenario.
Repeating the steps from our previous discussion and , we see that the gaps still open/close for $M=\pm3\sqrt{3}\lambda \sin\phi$ [\cref{eq:gapClosingMasses}], at momenta $\mathbf{k} = \{\mathbf{K}',\mathbf{K}\}$ and energies determined by $E - A(E) = -3 \lambda \cos\phi$.

In this "proof-of-principle" model, we have manually added the Semenoff mass and Haldane term to the $A$ and $B$ sites. We will next consider scenarios where these terms result from the ISR itself and correspond to what we call "latent" Semenoff and Haldane masses.

\subsection{Latent Semenoff mass}\label{Sec: Latent Semenoff}
\begin{figure}[!hbt]
    \centering
    \includegraphics[]{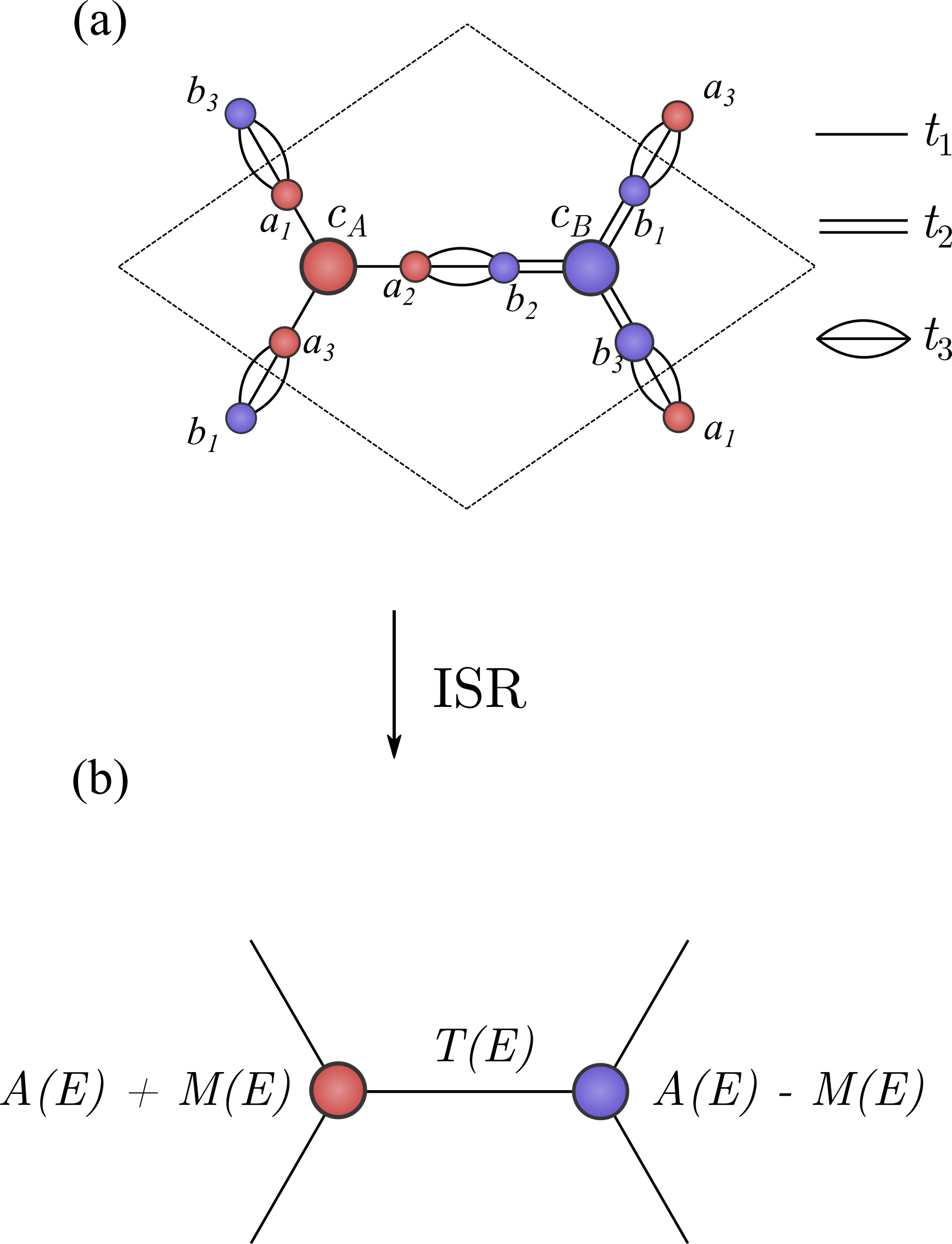}
    \caption{(a) Modified graphyne model and (b) ISR to graphene on sites $A$ and $B$, which are marked by larger circles than the rest. This now results in a latent Semenoff mass $M(E)$.}
    \label{fig:modifiedgraphyneLattice}
\end{figure}
\begin{figure}[!hbt]
    \centering
    \includegraphics[width=\columnwidth]{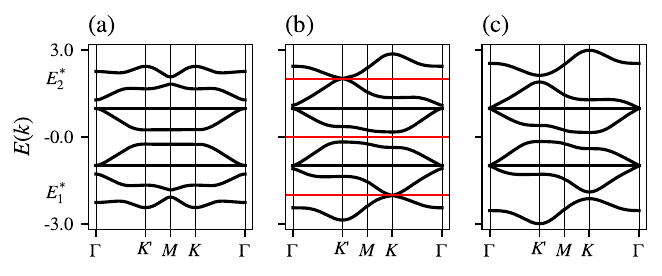}
    \caption{(a)-(c) Band structures of $\alpha$-graphyne, for unequal values of the latent Haldane mass. The bands cross the energies $E_1^*$ and $E_2^*$, predicted from \cref{Eq: Dirac Cone energies Lat Sem}. Parameter choices $(t_3,t,\delta)$, defined below \cref{Eq: latent Semenoff mass term}, are (a) $(1,1,0)$, (b) $(1,1,\delta_c\approx0.271558)$, and (c) $(1,1,0.35)$. We also use $(\lambda,\phi) = (0.2,\pi/2)$ in all cases. The red lines in (b) correspond to solutions $E^*$ to $E^* - \mathcal{A}(E^*) = 0$.}
    \label{fig: graphyne bandstructures}
\end{figure}

We start by modifying the $\alpha$-graphyne lattice.
As we see in the following, the key is to introduce different ``hopping neighborhoods'' to sites A and B, respectively.
This results in a latent Semenoff mass that becomes apparent after performing the ISR. 
The principle of different hopping neighborhoods is illustrated in \cref{fig:modifiedgraphyneLattice}. The bonds connecting red with red have strength $t_1$, blue with blue $t_2$, and red with blue $t_3$. The ISR is now given by 

\begin{equation*}
H_E(\mathbf{k}) = \begin{pmatrix}
    \frac{3Et_1^2}{E^2-t_3^2} & \frac{t_1t_2t_3}{E^2-t_3^2}g(\mathbf{k}) \\
    \frac{t_1t_2t_3}{E^2-t_3^2}g^*(\mathbf{k}) & \frac{3Et_2^2}{E^2-t_3^2}
\end{pmatrix},
\end{equation*}
which may be written as
\begin{align*}
    H_E(\mathbf{k}) & = \begin{pmatrix}
        \mathcal{A}(E) + M(E) & \frac{t_1t_2t_3}{E^2-t_3^2}g(\mathbf{k}) \\
        \frac{t_1t_2t_3}{E^2-t_3^2}g^*(\mathbf{k}) & \mathcal{A}(E) - M(E)
    \end{pmatrix}. 
\end{align*}  
We identified the energy dependent onsite term $\mathcal{A}(E)$ and the emergent energy dependent Semenoff mass $M(E)$, given by
\begin{equation}\label{Eq: latent Semenoff mass term}
\begin{split}
    \mathcal{A}(E) &\equiv \frac{3E(t_1^2+t_2^2)}{2(E^2-t_3^2)}\\
    M(E) &\equiv \frac{3E(t_2^2 - t_1^2)}{2(E^2-t_3^2)}.    
\end{split}
\end{equation}
For simplicity, we now set $t_1 = t+\delta$ and $t_2 = t-\delta$; the above two equations then become
\begin{align}
    \mathcal{A}(E) &\equiv \frac{3E(t^2+\delta^2)}{(E^2-t_3^2)} \nonumber\\
    M(E) &\equiv -\frac{6Et\delta}{(E^2-t_3^2)} \label{eq:LatentSemenoffMassTerm}\,.
\end{align}

Once again, we can add the TR-symmetry breaking term (Haldane mass) by hand, connecting the sites where we perform the ISR.
This results in the reduced Bloch-Hamiltonian
\begin{equation*}
    H_E(\mathbf{k}) + \begin{pmatrix}
        \lambda f_{\phi}(\mathbf{k}) & 0 \\
        0 & \lambda f_{-\phi}(\mathbf{k})
    \end{pmatrix} \,.
\end{equation*}
Note that in this setup, inversion symmetry has been broken by the presence of the third hopping, which is why it induces the mass term in the ISR picture. The measure of inversion-symmetry breaking is given by the parameter $\delta$. 

\begin{figure}[!t]
    \centering
    \includegraphics[width=\columnwidth]{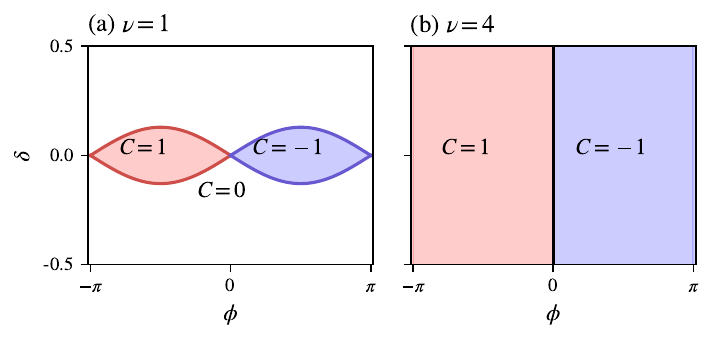}
        \caption{Phase diagram of the modified graphyne model (a) for $\nu=1$ and (b) $\nu=4$. The fixed parameters are $t=t_3=1$ and $\lambda=0.2$.}
        \label{fig: Phase Diag Latent Semenoff}
\end{figure}

\begin{figure}[!t]
    \centering
    \includegraphics[width=\columnwidth]{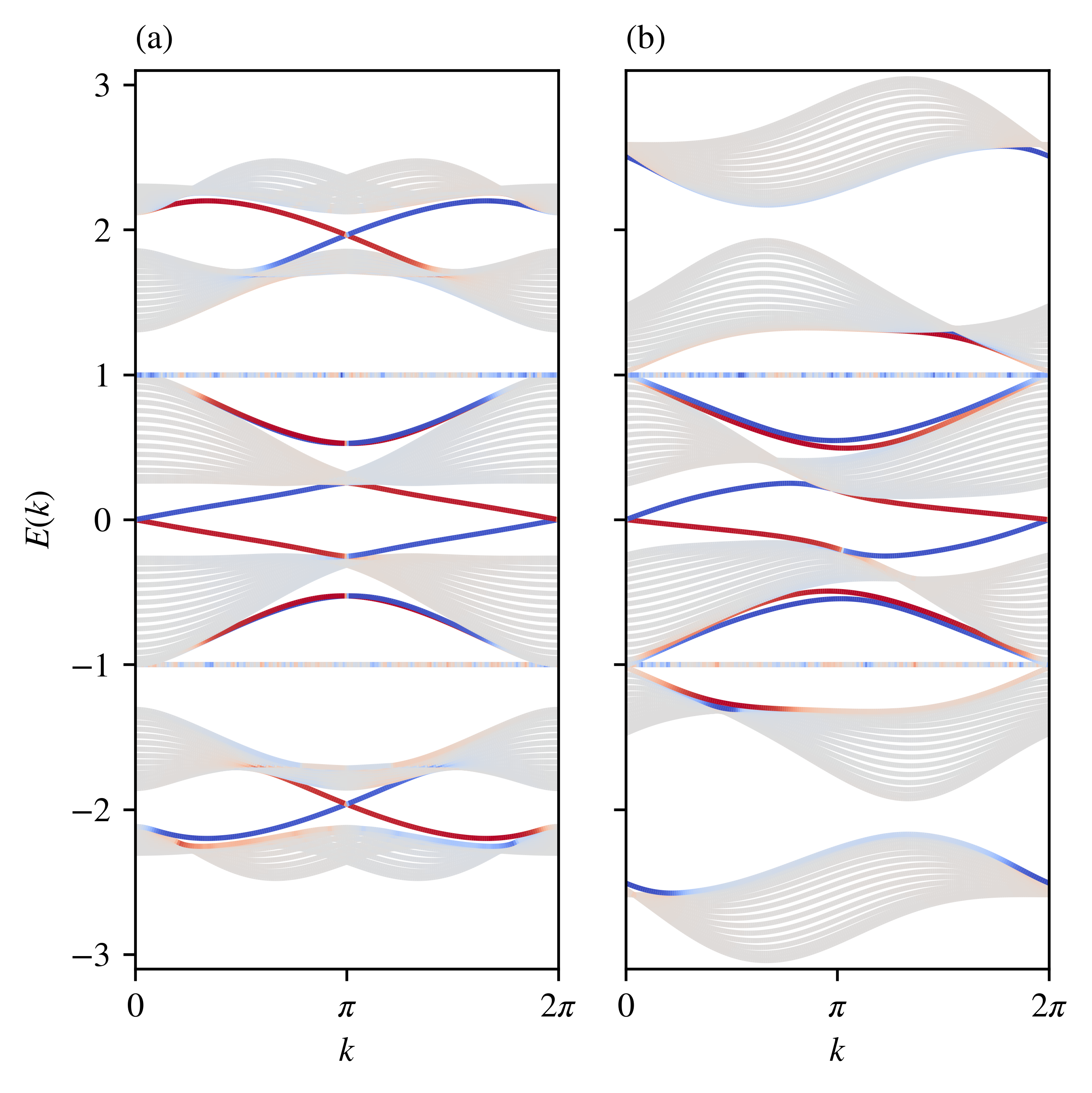}
    \caption{Bandstructure of the latent Semenoff mass model in a ribbon geometry, corresponding to the band structures and parameter choices $(t_3,t,\delta)$ given by (a) $(1,1,0)$ and (b) $(1,1,0.35)$. We also use $(g,\phi) = (0.2,\pi/2)$ in both cases. There are two flat bands at $E(k) = \pm 1$. For fillings $\nu=1$ and $\nu=7$, there is (a) a topological phase and (b) a trivial phase.
    Notice how the $\nu=4$ filling always shows topological edge modes, in agreement with the phase diagram in \cref{fig: Phase Diag Latent Semenoff} (b).}
    \label{fig: graphyne_Ribbons}
\end{figure}

For the original Haldane model, the band structure becomes gapless at $M = +3\sqrt{3}\lambda \sin{\phi}$ (with $\mathbf{k} = \mathbf{K}'$), and at $M = -3\sqrt{3}\lambda \sin{\phi}$ (with $\mathbf{k} = \mathbf{K}$) [cf. \cref{eq:gapClosingMasses}].
For our model with a latent Semenoff mass, one only has to replace $M$ by $M(E)$.
Note that the energy $E$ in this expression is still a free parameter; the allowed values $E^*$ are the solutions to $E^* - \mathcal{A}(E^*) = 0$ \footnote{This equation can be derived in a similar manner as in the latent Haldane model discussed in \cref{Sec: proof of principle}.}.
For our specific system, there are three solutions to the latter equation; they read
\begin{equation}\label{Eq: Dirac Cone energies Lat Sem}
    E^{*}_0 = 0 \quad \text{and} \quad E^{*}_\text{1,2} = \pm \sqrt{3(t^2+\delta^2)+t_3^2}.
\end{equation}
Inserting these values into $M(E^*) = \pm 3\sqrt{3}\lambda \sin{\phi}$ and using \cref{eq:LatentSemenoffMassTerm}, we obtain three equations that tell us for which parameter values the spectrum becomes gapless.
Note that, by analogy to the original Haldane model, these points also correspond to phase transitions.
As an example, we take $t_3 = 1$, $t=1$, $\lambda = 0.2$ and $\phi = \pi/2$, with varying $\delta$.
For the given model (with non-vanishing $\lambda \sin\phi$), the gap at $E=0$ cannot be closed by the energy-dependent Semenoff mass, as $M(0) = 0$; however, the other gaps can.
The critical value of $\delta$ can be found from $M(E_{1,2}^*) = \pm 3\sqrt{3}\lambda \sin{\phi}$, and it is found to be $\delta_c = \pm 0.271558$. This result is confirmed by the band structures shown in \cref{fig: graphyne bandstructures}: the topological phase with a nonzero latent Semenoff mass in \cref{fig: graphyne bandstructures}(a), the gapless band structure at $\delta_c$ in \cref{fig: graphyne bandstructures}(b), and the trivial phase in \cref{fig: graphyne bandstructures}(c).
Besides this specific example, we can construct a phase diagram for explicit parameter values.
For instance, we find that if $\nu=1$ bands are filled, the critical line (corresponding to the gap between the lowest and second-lowest band) is given by 
\begin{equation}
    \delta_c = \pm \sqrt{\frac{-3t^2-t_3^2+\sqrt{(3t^2+t_3^2)^2+81\lambda^2\sin^2\phi}}{6}},
\end{equation}
while for $\nu=4$ filled bands, the two phases are separated by the vertical line $\phi=0$. This allows us to construct the phase diagrams shown in Figs.~\ref{fig: Phase Diag Latent Semenoff}(a) and \ref{fig: Phase Diag Latent Semenoff}(b) for fillings $\nu=1$ and $\nu=4$, respectively.
This result confirms the expectation from the bulk-boundary correspondence and agree with \cref{fig: graphyne_Ribbons}, where topological edge states appear when $|\delta| < |\delta_c|$ for filling $\nu=1$ (and also $\nu=7$) [see \cref{fig: graphyne_Ribbons}(a)], but not in \cref{fig: graphyne_Ribbons}(b), where $\delta>\delta_c$.
Additionally, the $\nu=4$ filling always showcases topological edge modes, independently of the value of $\delta$. 

\subsection{Latent Haldane mass}\label{Sec: Latent Hald mass}
\begin{figure}[!hbt]
    \centering
    \includegraphics[]{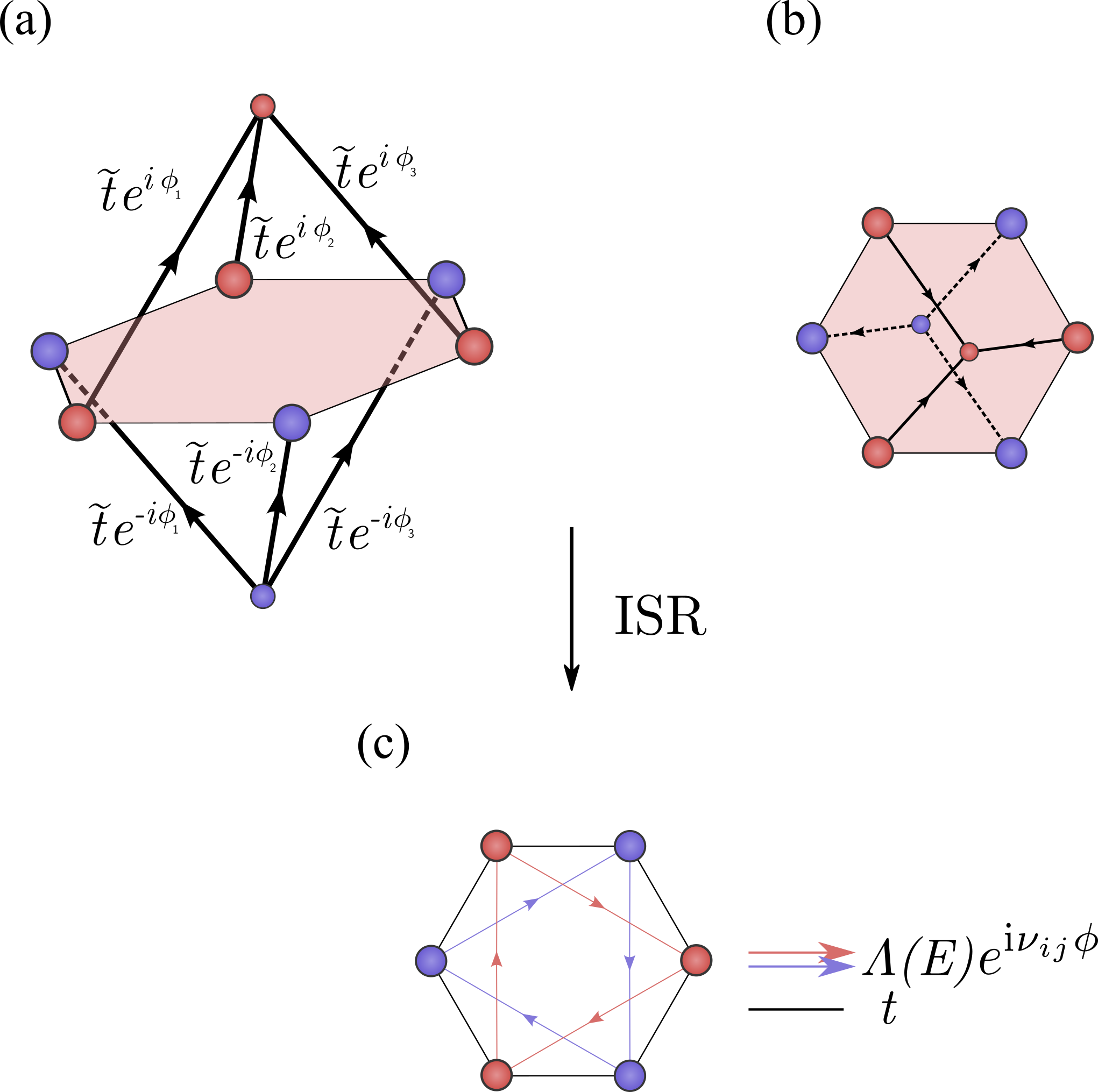}
    \caption{Cell of a decorated honeycomb lattice. a)  Side view and b) top view. c) ISR onto sites $A$ and $B$, resulting in the Haldane model, with a latent complex Haldane coupling mass $\Lambda(E)$.}
    \label{fig:LatentHaldaneMasslattice}
\end{figure}

We now investigate a decorated lattice that exhibits a latent Haldane mass. The initial setup is sketched in \cref{fig:LatentHaldaneMasslattice}, where a hexagonal plaquette is connected to two additional sites through complex hoppings $\Tilde{t}e^{\pm i\phi_m}$. \cref{fig:LatentHaldaneMasslattice}(a) shows a side view, with the two sites being located on top and bottom of the plaquette, \cref{fig:LatentHaldaneMasslattice}(b) displays a top view of the same setup. In \cref{fig:LatentHaldaneMasslattice}(c), the ISR to the six sites surrounding the plaquette is shown.
We allow for three different hopping phases $\phi_m$, $m=1,2,3$; otherwise, the system is always in a topologically trivial phase. The hoppings must also always have a relative phase difference of $2\pi/3$ to generate no net magnetic flux on the plaquette.
Using these constraints on the phase, and constructing a lattice from the unit cell, the ISR yields the following energy-dependent $2\times2$ Bloch Hamiltonian,
\begin{equation}
    H_E(\mathbf{k})=\begin{pmatrix}
    A(E) + \Lambda(E) f_{\frac{2\pi}{3}}(\mathbf{k}) & tg(\mathbf{k}) \\
        tg^*(\mathbf{k})& A(E) + \Lambda(E) f_{-\frac{2\pi}{3}}(\mathbf{k}),
    \end{pmatrix}
\end{equation}
where $A(E)=\Lambda(E)/3=\Tilde{t}^2/E$, and $g(\mathbf{k})$ and $f_\phi(\mathbf{k})$ are given below \cref{Eq: Haldane Bloch Hamiltonian}. 

\begin{figure}[!hbt]
    \centering
    \includegraphics[width=\columnwidth]{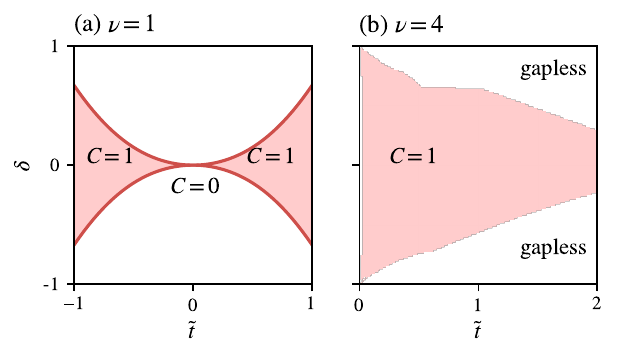}
    \caption{Phase diagram of the full latent Haldane model for (a) $\nu=1$ filling, obtained analytically from \cref{Eq: LH phase diag nu=1} and (b) $\nu=4$ filling, obtained numerically. The free parameters were set to $t=t_3=1$.}
    \label{fig: LAtent Full Haldane Phase diagram}
\end{figure}

In order to make the setup more interesting, we should also include the previous decorations introduced in \cref{fig:modifiedgraphyneLattice}. This results in an additional latent Semenoff mass given by \cref{Eq: latent Semenoff mass term}. The bandstructure of this system is shown in \cref{fig: Full Haldane bandstructures}, as the parameters are varied across a topological phase transition. Figure \ref{fig: Full Haldane bandstructures}(b) shows the band structure at the transition point, with the Dirac cones at $K$ for the highest and lowest gaps, while Figs.~\ref{fig: Full Haldane bandstructures}(a) and \ref{fig: Full Haldane bandstructures}(c) show, respectively, the bandstructure in the topological and trivial phases, for the same gaps.
Once again, we can find the energies where, for suitable choice of the other parameters, gaps close, by an analog procedure as done in the previous examples, though now taking into account that both the Semenoff mass and the Haldane mass are energy-dependent. The energies where gaps may close then have to satisfy 
\begin{equation}\label{Eq: Mgap Full Haldane}
    E-\frac{3E(t^2+\delta^2)}{(E^2-t_3^2)}-\frac{9\Tilde{t}^2}{2E} = 0.
\end{equation} 
The solutions of this equation are represented in \cref{fig: Full Haldane bandstructures}(b) by red lines.
When the energies corresponding to the fillings $\nu=1$ and $\nu=9$ are plugged in \cref{eq:continuumHallConductivity}, the following gap closing condition induced by a topological phase transition is obtained: 
\begin{figure}[!t]
    \centering
    \includegraphics[width=\columnwidth]{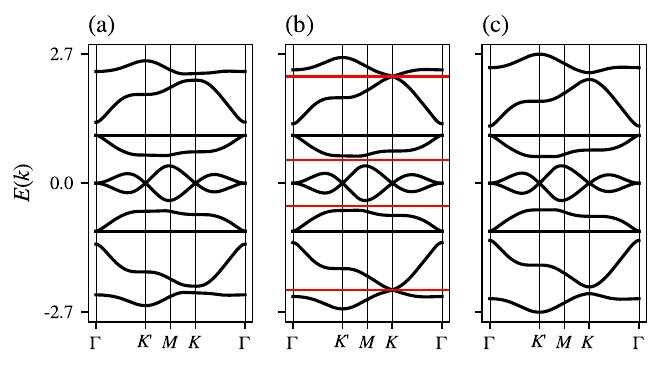}
    \caption{Band structures of the full latent Haldane model. The bands cross at the energies predicted from \cref{Eq: Mgap Full Haldane}. Parameter choices $(t_3,t,\delta)$ are (a) $(1,1,0.1)$, (b) $(1,1,\delta_c\approx0.14441)$, and (c) $(1,1,0.2)$.
    We have used $\Tilde{t} = 0.5$ in all cases.}
    \label{fig: Full Haldane bandstructures}
\end{figure}
\begin{equation}\label{Eq: LH phase diag nu=1}
\begin{split}
        &\left(6 \delta \pm9 \Tilde{t}^2\right) \left(\delta ^2+t^2\right)+ \\ &\delta \sqrt{\left(6 \delta ^2+9 \Tilde{t}^2+6 t^2+2 t_3^2\right)^2-72 \Tilde{t}^2
   t_3^2} +\delta  \left(2 t_3^2-9 \Tilde{t}^2\right)=0.
\end{split}
\end{equation}
With these two curves, we can then construct the phase diagram shown in \cref{fig: LAtent Full Haldane Phase diagram}(a), in terms of the parameters $\delta$ and $g$ (we have set $t=t_2=1$ in the figure) for the $\nu=1$ gap. In \cref{fig: LAtent Full Haldane Phase diagram}(b), the phase diagram for the $\nu=4$ gap is shown.
It was calculated numerically because \cref{Eq: Mgap Full Haldane} no longer applies; the gap closure occurs away from high-symmetry points, rendering the equation ineffective.

\begin{figure}[]
    \centering
    \includegraphics[width=\columnwidth]{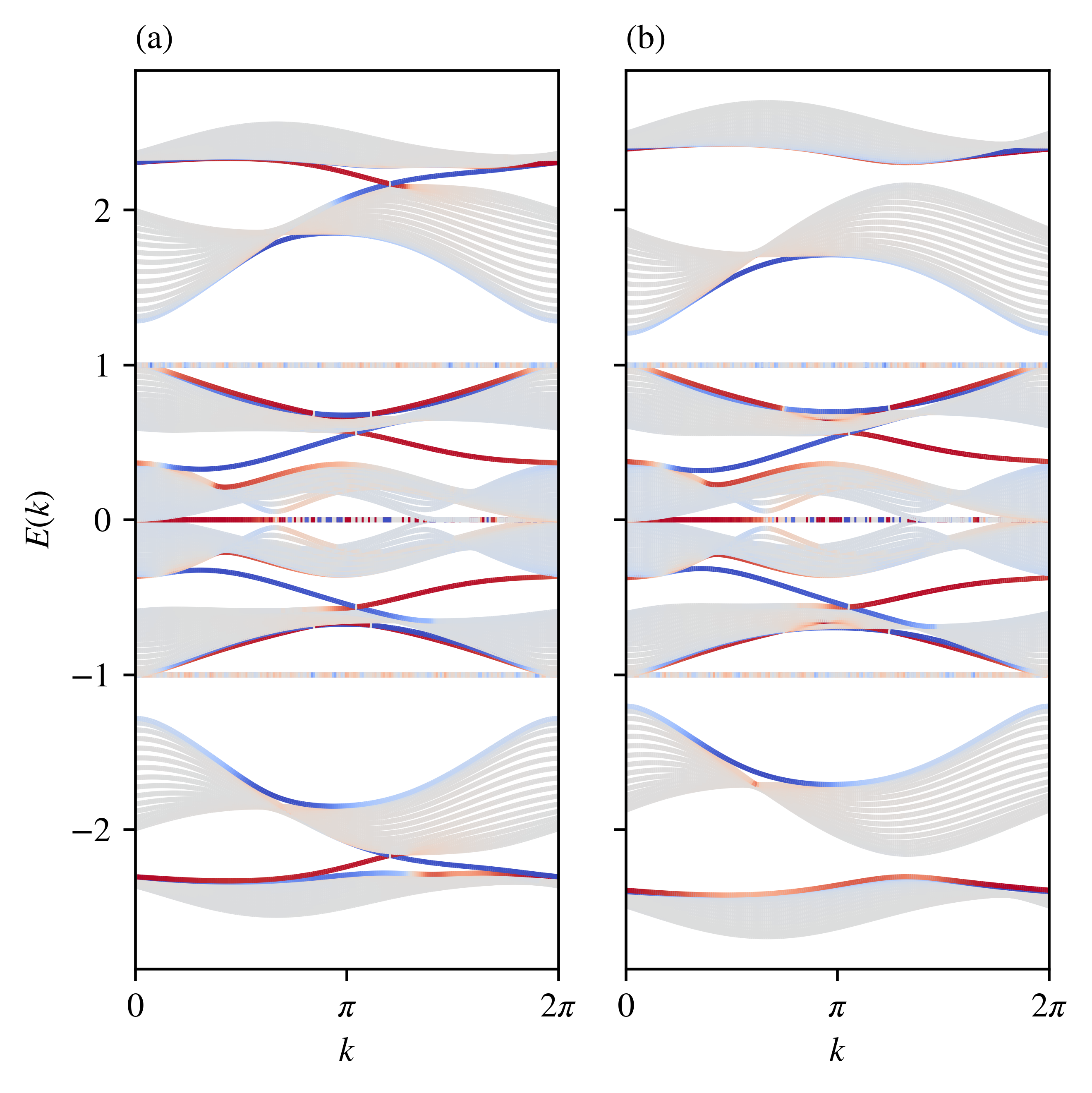}
    \caption{Bandstructure of the latent Semenoff mass model, in a ribbon geometry, corresponding to the band structures and parameter choices in Figs.~\ref{fig: Full Haldane bandstructures}(a) and (c). In all cases, we have used $\Tilde{t} = 0.5$. There is (a) a topological phase for the $\nu=1$ and $\nu=9$ fillings when $(t_3,t,\delta)=(1,1,0.1)$, and (b) a trivial phase for those same fillings when $(t_3,t,\delta)=(1,1,0.2)$. Notice how the $\nu=4$ filling always showcases topological edge modes, in agreement with the phase diagram shown in \cref{fig: LAtent Full Haldane Phase diagram} (b).}
    \label{fig: FullHaldane_Ribbons}
\end{figure}

In \cref{fig: FullHaldane_Ribbons}, we plot the spectrum in the ribbon geometry and indeed observe topological edge states appearing, as predicted by the phase diagram constructed in \cref{fig: LAtent Full Haldane Phase diagram}.

\section{Generalizations}\label{sec: Generalizations}

\begin{figure}[!hbt]
	\begin{center}
		\def\svgwidth{0.9\linewidth}
		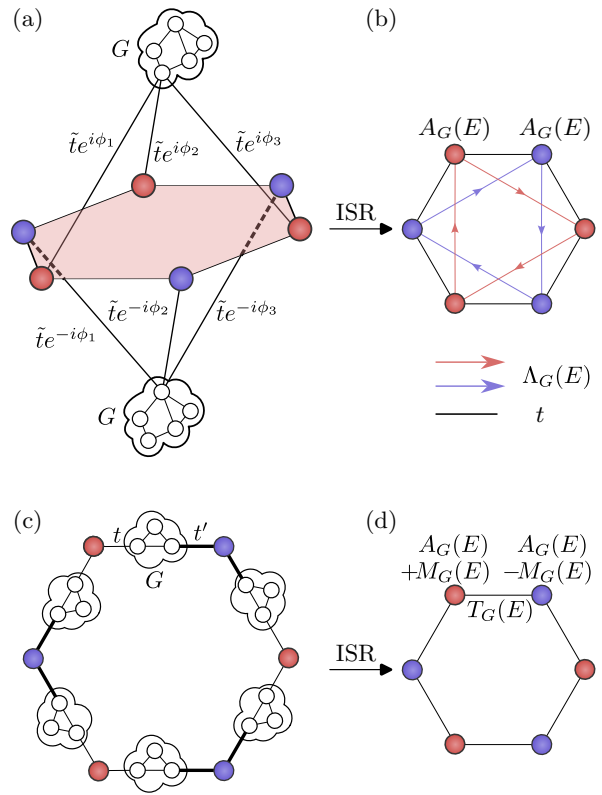
		\caption{Cell of a decorated hexagonal plaquette. (a), (c) Side view. (b), (d) ISR onto sites $A$ and $B$, resulting in the Haldane model with (b) a latent complex Haldane coupling mass $\Lambda_G(E)$, and (d) a latent Semenoff mass $A_G(E) \pm M_G(E)$. Arrows denote the directions of complex-valued couplings.
        }
		\label{fig:generalizationPrinciple}
	\end{center}
\end{figure}

Let us now briefly discuss how the above constructions of latent Haldane models can be generalized.
We start by generalizing the principle behind the latent Haldane mass term, of which we have realized a simple version in \cref{fig:LatentHaldaneMasslattice}.
There, we coupled all the $A$ sites via complex-valued hoppings to a single site and all the $B$ sites, also via complex-valued hoppings, to another single site; cf. \cref{fig:LatentHaldaneMasslattice}(a).
It can be shown (see \cref{App:ConstructionPrincniples}) that one can replace each of these two single sites with the same arbitrary substructure $G$; cf. \cref{fig:generalizationPrinciple}(a).
Indeed, as long as only a single site in this substructure is coupled to the $A$ (or $B$) sites, the ISR of this setup features the structure depicted in \cref{fig:generalizationPrinciple}(b), where the functional form of $A_G(E)$ and $\Lambda_G(E)$ depends on the choice of the graph $G$. As a consequence, its Bloch-Hamiltonian will feature an energy-dependent Haldane mass.

Next, let us consider the latent Semenoff term.
In \cref{fig:modifiedgraphyneLattice}, a simple realization of this term is shown.
There, we replaced the coupling between the two sites $A$ and $B$ by a chain of two sites $a_2$ and $b_2$, and then coupled this chain asymmetrically to $A$ and $B$.
There are many possible generalizations, but perhaps the simplest one is to replace the chain of two sites with a general reflection-symmetric structure $G$; see \cref{fig:generalizationPrinciple}(c).
Upon performing the ISR onto the $A$ and $B$ sites, energy-dependent on-site terms $A_G(E) \pm M_G(E)$, of which the details depend on the graph $G$, appear on the $A$ and $B$ sites, see \cref{fig:generalizationPrinciple}(d).

Before concluding, let us remark that one could combine these two principles to obtain a latent Haldane model featuring both a Haldane and a Semenoff mass term.

\section{Conclusion} \label{Sec: Conclusion}

The Haldane model has been foundational in advancing the theoretical understanding of topological insulators. By capturing essential properties, the Haldane model facilitates the exploration of phenomena that arise in more complex and realistic topological materials. 
Building upon the idea of reducing complicated lattice structures to paradigmatic models \cite{Eek2024EmergentModels}, we have developed families of lattice structures that, through the application of an ISR, produce energy-dependent Haldane models. This approach allows us to access the features of the Haldane model to illuminate the behavior of these intricate systems, offering insights that would be challenging to obtain through a direct analysis of the complete and complicated structures.
For instance, this framework permits the construction of a phase diagram by direct analytic calculations, instead of relying on numerical computations. This idea was first shown to yield useful insights in the case of $\alpha$-graphyne, where the gap-closing energies can be calculated using the ISR.
Hereafter, we have demonstrated that applying the ISR to various decorated lattice models can produce a latent, energy-dependent Semenoff mass, which breaks latent reflection symmetry without necessitating an on-site staggered potential. Notably, this approach enables topological phase transitions through the modulation of hopping parameters, provided that an additional complex hopping term is introduced. The resulting energy-dependent Haldane model was subsequently used to analytically derive a phase diagram for the specific fillings that support these distinct phases.
This was followed by a construction of a lattice that yielded latent Haldane masses arising from nearest-neighbor complex hoppings, with a $2\pi/3$ phase difference. After following a similar procedure, a phase diagram was analytically constructed for one filling, while it had to be numerically calculated for another. 

With the above ingredients, the construction principle can be generalized to a broader family of models that exhibit similar characteristics. While these systems might initially appear complex due to the large number of bands in their original formulation, utilizing the non-linear Haldane-Bloch Hamiltonian enables the identification of energy gaps that host topological states and of the precise points at which phase transitions occur.

The work represented here offers a starting point for further development towards the construction of more complicated structures that reduce to other paradigmatic models. A logical first extension would be to incorporate spin to construct generalized Kane-Mele type models. Furthermore, one could consider structures in three spatial dimensions. This offers many more crystalline symmetries and may, consequently, host novel topological phases that exist only by virtue of (latent) crystalline symmetries.

\begin{acknowledgments}
    A.M. and C.M.S. acknowledge the project TOPCORE with project number OCENW.GROOT.2019.048 which is financed by the Dutch Research Council (NWO). L.E. and C.M.S. acknowledge the research program “Materials for the Quantum Age” (QuMat) for financial support. This program (registration number 024.005.006) is part of the Gravitation program financed by the Dutch Ministry of Education, Culture and Science (OCW). 
    M.R. acknowledges fruitful discussions with G. E. Sommer.
\end{acknowledgments}

 \bibliography{Refs,Bibtex_Final}

\appendix
\onecolumngrid

\section{Details on the construction principles for a latent }\label{App:ConstructionPrincniples}
\begin{figure}[!hbt]
	\begin{center}
		\def\svgwidth{0.5\linewidth}
		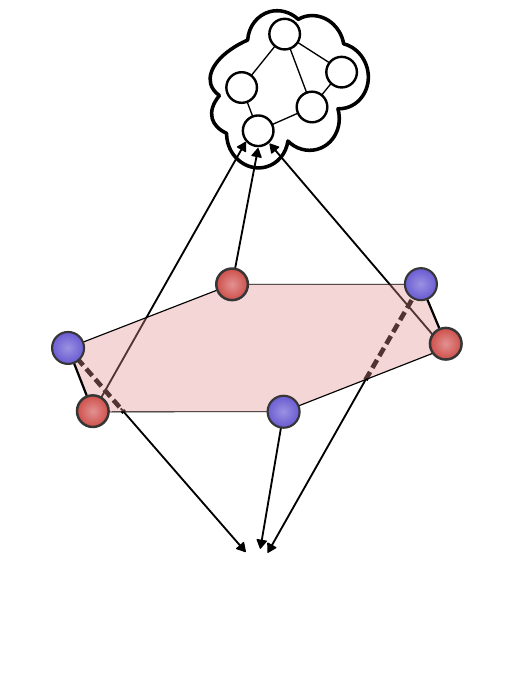
		\caption{As in \cref{fig:generalizationPrinciple}(a), though with enumerated sites.
        }
		\label{fig:generalizationPrincipleAppendix}
	\end{center}
\end{figure}

In \cref{sec: Generalizations} of the main text, we have discussed general construction principles for systems with a latent Haldane mass or a latent Semenoff mass.
Here, we discuss the underlying mathematical details for these construction principles.

\subsection{Latent Haldane mass}
Let us start by the principle for generating a latent Haldane mass, as depicted in \cref{fig:generalizationPrinciple}(a).
This figure is again depicted in \cref{fig:generalizationPrincipleAppendix}(a), though with additional numbers imprinted on each site to ease the discussion.
The central six sites, that is, the red sites $1$ to $3$, and the blue sites $4$ to $6$ form the set $S$ over which we perform the ISR.
The remaining sites $7$ to $11$ (upper network $G$) and $12$ to $16$ (lower network $G$) form the set $\sbar$.
With this enumration of sites, we can analyze the ISR 
\begin{equation}
    \ISR{S}{\en} = \ham_{SS} - \ham_{\ssb} \left(\ham_{\sbb} - \en{}\, I \right)^{-1} \ham_{\sbs} \,,
\end{equation}
of the depicted network.
It is helpful to analyze the terms individually.
The first, $\ham_{SS}$ is just the $6 \times{} 6$ matrix describing the isolated plaquette.
It is the second term where things get interesting.
The matrices
\begin{equation}
    \ham_{\ssb} = \tilde{t} \begin{pmatrix}
        1 & 0 & 0 & 0 & 0 & 0 & 0 & 0 & 0 & 0 \\
        e^{2\pi i /3} & 0 & 0 & 0 & 0 & 0 & 0 & 0 & 0 & 0 \\
        e^{4\pi i /3} & 0 & 0 & 0 & 0 & 0 & 0 & 0 & 0 & 0 \\
        0 & 0 & 0 & 0 & 0 & 1 & 0 & 0 & 0 & 0  \\
        0 & 0 & 0 & 0 & 0 & e^{2\pi i /3} & 0 & 0 & 0 & 0  \\
        0 & 0 & 0 & 0 & 0 & e^{4\pi i /3} & 0 & 0 & 0 & 0 
    \end{pmatrix}
\end{equation}
and $\ham_{\sbs} = \left( \ham_{\ssb}\right)^\dagger$, where $\dagger$ denotes the hermitian conjugate, are sparse, since only a single site in each of the two copies of the network $G$ is coupled to the sites in $S$.
Finally,
\begin{equation}
    \left(\ham_{\sbb} - \en{}\, I \right)^{-1} = \begin{pmatrix}
        \mathbf{g}(E) & \mathbf{0} \\
        \mathbf{0} & \mathbf{g}(E)
    \end{pmatrix}
\end{equation}
where both $\mathbf{g}(E)$ and $\mathbf{0}$ is a $5 \times 5$ matrix; the former is the expression $\left(\ham_G - \en{}\, I \right)^{-1}$ (with $\ham_G$ denoting the Hamiltonian describing the isolated network $G$), and the latter is the $5 \times 5$ matrix of all zeros.

The crucial point is to realize that, due to the special structure of those matrices, the second term becomes equal to 
\begin{equation} \label{eq:complexCouplingsISR}
    \tilde{t} \, g_{11}(E)
    \begin{pmatrix}
 1 & e^{-\frac{2 i \pi }{3}} & e^{\frac{2 i \pi }{3}} & 0 & 0 & 0 \\
 e^{\frac{2 i \pi }{3}} & 1 & e^{-\frac{2 i \pi }{3}} & 0 & 0 & 0 \\
 e^{-\frac{2 i \pi }{3}} & e^{\frac{2 i \pi }{3}} & 1 & 0 & 0 & 0 \\
 0 & 0 & 0 & 1 & e^{-\frac{2 i \pi }{3}} & e^{\frac{2 i \pi }{3}} \\
 0 & 0 & 0 & e^{\frac{2 i \pi }{3}} & 1 & e^{-\frac{2 i \pi }{3}} \\
 0 & 0 & 0 & e^{-\frac{2 i \pi }{3}} & e^{\frac{2 i \pi }{3}} & 1
\end{pmatrix} \,,
\end{equation}
where $g_{11}$ denotes a matrix element of $\mathbf{g}(E)$.
Identifying $\tilde{t} \, g_{11}(E)$ with $A_G(E)$, and $\tilde{t} \, g_{11}(E) e^{-\frac{2 i \pi }{3}}$ with $\Lambda_G(E)$, and taking into account the first term $\ham_{SS}$, we see that the ISR takes exactly the form as visualized in \cref{fig:generalizationPrinciple}(b).

After this extensive calculation, it becomes clear that the structure of \cref{eq:complexCouplingsISR} is independent of how exactly the network $G$ looks like; as long as only a single site in $G$ is coupled to the red (or blue) sites in the plaquette, the second term of the ISR will always be given by \cref{eq:complexCouplingsISR}.
Moreover, since the first term $\ham_{SS}$ in the ISR is independent of $G$, we see that one can replace $G$ by an arbitrary network, as claimed in the main text.

\subsection{Latent Semenoff mass}
\begin{figure}[!hbt]
	\begin{center}
		\def\svgwidth{0.5\linewidth}
		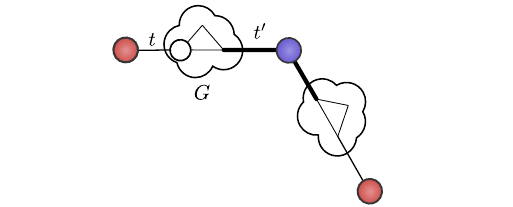
		\caption{Half of the plaquette depicted in \cref{fig:generalizationPrinciple}(c), and with enumerated sites.
        }
		\label{fig:generalizationPrincipleSemenoffAppendix}
	\end{center}
\end{figure}
Finally, we come to the construction principle for a latent Semenoff mass.
A simple setup for understanding the details of the underlying principle is given in \cref{fig:generalizationPrincipleSemenoffAppendix}.
It shows a part of the plaquette depicted in \cref{fig:generalizationPrinciple}(c), with numbers added to the sites to ease the discussion.
We reduce over the two red ($1$ and $3$) and the blue ($2$) sites, which thus form the set $S$.
The sites $4$ to $9$ thus form the set $\sbar$.
In this case, since the sites in $S$ are not connected to each other, and also have no on-site potential, the matrix $\ham_{SS}$ occurring in the ISR just vanishes.
In the second term of the ISR, we have the matrices
\begin{equation}
    \ham_{\ssb} = \tilde{t}
    \begin{pmatrix}
        t & 0 & 0 & 0 & 0 & 0 \\
        0 & t' & 0 & t' & 0 & 0 \\
        0 & 0 & 0 & 0 & t & 0
        \end{pmatrix} \,,
\end{equation}
$\ham_{\sbs} = \left( \ham_{\ssb}\right)^\dagger$, as well as
\begin{equation}
    \left(\ham_{\sbb} - \en{}\, I \right)^{-1} = 
    \begin{pmatrix}
        \mathbf{g}(E) & \mathbf{0} \\
        \mathbf{0} & \mathbf{g}(E)
    \end{pmatrix}
\end{equation}
with
\begin{equation}
    \mathbf{g}(E) := \begin{pmatrix}
        g_{11} & g_{12} & g_{13} \\
        g_{12} & g_{22} & g_{23} \\
        g_{13} & g_{23} & g_{33}
    \end{pmatrix}(E) \,,
\end{equation}
where we have assumed that $\ham_{\sbb}$ is completely real-valued.
$\mathbf{0}$ denotes the $3\times{}3$ matrix with all entries zero.

The crucial point is to realize that, since $G$ is reflection symmetric, the two matrix elements $g_{11}$ and $g_{22}$ are identical \footnote{To see this fact, one has to write the permutation matrix $M$ corresponding to the reflection symmetry of $G$.
$M$ permutes the two sites in $G$ that are connected to $S$.
From the commutation of $M$ and $\ham_G$ (the Hamiltonian describing $G$), it follows that the corresponding diagonal entries of $\ham_G$ must be identical.
Since $M$ commutes also with $\left(\ham_G - E I \right)^{-1}$, the equality of the two diagonal elements holds there as well.}; we denote them by $g_d$ ($d$ for diagonal).
Furthermore, we note that this is always the case provided that $G$ is reflection-symmetric and that the sites in $G$ coupled to $S$ are those that are mirror-images of each other.
As a result of the equality of the , the ISR becomes
\begin{equation}
\begin{pmatrix}
        g_d(E)\, t^2 & t\, t' \,g_{12}(E) & 0 \\
 t\, t' g_{12}(E) & 2 t'^2 g_d(E)  & t t' g_{12}(E) \\
 0 & t \,t' g_{12}(E) & g_d(E)\, t^2
\end{pmatrix}\,.
\end{equation}
We note that the prefactor $2$ in the second diagonal element comes from the fact that the site $2$ is coupled to two networks $G$, while the sites $1$ and $3$ are each coupled only to one network $G$.
When taking the ISR over the full plaquette and not only on the small part of it that we discussed here, each diagonal element of the ISR has a prefactor of $2$.
Specifically, the ISR over the full plaquette would be given by
\begin{equation}
\begin{pmatrix}
a(E) & T_G(E) & 0 & 0 & 0 & T_G(E)\\
T_G(E) & b(E)  & T_G(E) & 0 & 0 &0\\
0 & T_G(E) & a(E) & T_G(E) & 0 & 0\\
0 & 0 & T_G(E) & b(E) & T_G(E) & 0 \\
0 & 0 & 0 & T_G(E) & a(E) & T_G(E) \\
T_G(E) & 0 & 0 & 0 & T_G(E) & b(E)
\end{pmatrix}\,,
\end{equation}
where $T_G(E) = t\, t' g_{12}(E)$, $a(E) = 2  g_d(E)\, t^2$, and $b(E) = 2 t'^2 g_d(E)$.
If we further identity $A(E) := (t^2 + t'^2)g_d(E)$, and $M(E) := (t^2 - t'^2)g_d(E)$, we obtain exactly the ISR given in graphical form in \cref{fig:generalizationPrinciple}(d).


\end{document}